# Preliminary results of 3D-DDTC pixel detectors for the ATLAS upgrade

**Alessandro La Rosa** [a,1]
*CERN,*
*CH-1211 Geneve, Switzerland*
*E-mail:* alessandro.larosa@cern.ch

**M. Boscardin[b], G.-F. Dalla Betta[c], G. Darbo[d], C. Gemme[d], H. Pernegger[a], C. Piemonte[b], M. Povoli[c], S. Ronchin[b], A. Zoboli[c], N. Zorzi[b], E. Bolle[e], M. Borri[f], C. Da Via[g], S. Dong[h], S. Fazio[i], P. Grenier[h], S. Grinstein[l], H. Gjersdal[e], P. Hansson[h], F. Huegging[m], P. Jackson[h], M. Kocian[h], F. Rivero[f], O. Rohne[e], H. Sandaker[n], K. Sjobak[e], T. Slavicek[o], W. Tsung[m], D. Tsybychev[p], N. Wermes[m], and C. Young[h]**

[a] *CERN, CH-1211 Geneve, Switzerland*
[b] *Fondazione Bruno Kessler (FBK-irst), Via Sommarive 18, 38123 Povo di Trento, Italy*
[c] *University of Trento and INFN, Via Sommarive 14, 38123 Povo di Trento, Italy*
[d] *INFN – Genova, Via Dodecaneso 33, 16146 Genova, Italy*
[e] *University of Oslo, Fysisk Institutt, Postbooks 1048 Blindern, 0316 Oslo, Norway*
[f] *INFN and University of Torino, Via P. Giuria 1, 10125 Torino, Italy*
[g] *University of Manchester, Oxford Road, Manchester, M13 9PL, UK*
[h] *SLAC, 2575 Sand Hill Road, Menlo Park, CA 94025, USA*
[i] *University of Calabria, Via P. Bucci, Cubo 31 C, 87036 Arcavacata di Rende, Cosenza, Italy*
[l] *IFAE Barcelona, Edifici CN UAB Campus, 08193 Bellaterra (Barcelona), Spain*
[m] *University of Bonn, Nussallee 12, 53115 Bonn, Germany*
[n] *University of Bergen, Allegaten 55, 5007 Bergen, Norway*
[o] *Czech Technical University – Prague, Zlkova 4, 16636 Prague, Czech Republic*
[p] *Stony Brook, Stony Brook, 11794 NY, USA*

3D Silicon sensors fabricated at FBK-irst with the Double-side Double Type Column (DDTC) approach and columnar electrodes only partially etched through p-type substrates were tested in laboratory and in a 1.35 Tesla magnetic field with a $180 GeV$ pion beam at CERN SPS.

The substrate thickness of the sensors is about $200 \mu m$, and different column depths are available, with overlaps between junction columns (etched from the front side) and ohmic columns (etched from the back side) in the range from $110 \mu m$ to $150 \mu m$.

The devices under test were bump bonded to the ATLAS Pixel readout chip (FEI3) at SELEX SI (Rome, Italy). We report leakage current and noise measurements, results of functional tests with Am[241] $\gamma$-ray sources, charge collection tests with Sr[90] $\beta$-source and an overview of preliminary results from the CERN beam test.

---

[1] Speaker

## 1. Introduction

An upgrade of the LHC towards a 10 times higher luminosity will require tracking pixel detectors with unprecedented radiation tolerance [1]. Furthermore, the high track density will call for fast and high granularity pixel detectors with low radiation length and power consumption. Different types of solid-state sensors are being studied for the ATLAS upgrade of the tracking system within the ATLAS B-Layer replacement and Super-LHC programs.

The Silicon 3D sensors, which have been proposed and developed are potentially more radiation hard and have a faster charge collection than the standard planar sensors, owing to the innovative electrode configuration. Thus, these sensors represent an excellent candidate for the pixel sensors at ATLAS IBL and SLHC. In the paper we present the preliminary results of 3D Double-side Double Type Column fabricated at FBK-irst with the ATLAS Pixel readout chip.

## 2. 3D-DDTC Silicon Pixel Detectors

3D Double-side Double Type Column (DDTC) sensors have been fabricated at FBK-irst with a *non-passing-through* columns technology on p-type substrate [2,3]. Using this technology two batches of 3D DDTC sensors have been realized: one so-called 3D-DTC-2 (with the DRIE[1] performed at IBS (Fr)) and another one so-called 3D DTC-2B (with the DRIE performed in-house at FBK-irst). Figure 1 shows respectively a 3D-DDTC cross-section and a sketch of columnar electrodes in ATLAS pixels of 2E, 3E, and 4E type, whereas Table 1 summarizes the geometrical parameters of sensors from the two batches.

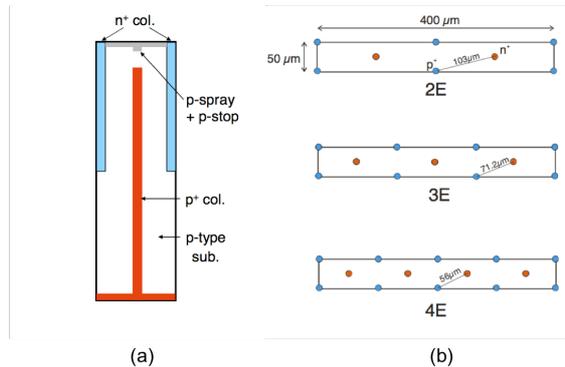

(a)                                         (b)

*Figure 1. (a) Sketch of 3D-DDTC cross-section; (b) Sketch of columnar electrodes in ATLAS pixels of 2E, 3E, and 4E type [2,3].*

A few samples of DDTC sensors with 2, 3 and 4 columns per pixel have been bump bonded and flip chipped to the ATLAS Pixel readout chip (FE-I3) [4]. The bump has been based on Indium and it has been carried out by SELEX [5].

The FE-I3 chip has been fabricated in a $0.25\mu m$ CMOS process and it consists of 2880 readout cells of $50\mu m$ x $400\mu m$ size arranged in a matrix of 18 x 160. The chip provides zero suppression in the readout with a typical threshold level of $3000 - 4000e$ and a noise of 200 - $230e$ as determined from fitting the threshold curve of the pixel (S-curve), when used with 3D-DDTC sensor. Analog information are obtained via the internal channel measurements of time-over-threshold (ToT) with an approximate resolution of 8 bit by unit of $25nsec$. The ToT-calibration is obtained for every pixel individually, by fitting the ToT response to input pulses injected over an internal capacitance [4].

---

[1] Deep Reactive Ion Etching

| Parameter | Unit | Value | |
|-----------|------|-------|-|
| | | **3D-DTC-2** | **3D-DTC-2B** |
| Substrate thickness | $\mu m$ | 200 | 200 |
| Junction column thickness | $\mu m$ | 100 -110 | 140 -170 |
| Ohmic column thickness | $\mu m$ | 180 -190 | 180 – 190 |
| Column overlap | $\mu m$ | 90 - 100 | 110 -150 |

*Table 1. Geometrical parameters of sensors from DTC-2 and DTC-2B batches.*

## 3. Lab characterization

Before including the assemblies in a beam test at CERN SPS, the detectors under tests (DUTs) have been studied by measuring: a) leakage currents, b) threshold and noise, c) response to γ and β radioactive sources.

An overview of IV curves of DDTC detectors from first batch (DTC-2) is shown in figure 2, where the breakdown occurs at a voltage of about -70 *V*. As has already been described in [3], the DTC-2B sensors present an early breakdown, which is related to the presence of local defects likely occurred during the assembly.  This problem is still under investigation.

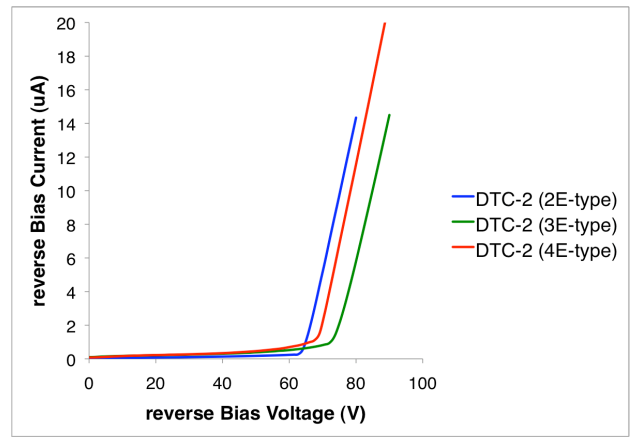

*Figure 2. Overview of IV curves of 3D-DDTC detectors from first batch (DTC-2).*

Threshold and noise measurements have been performed on each pixel, where only pixels with a charge deposit above the threshold are taken into account for readout by the front-end electronics. Table 2 summarizes the results of the threshold and noise measurements (performed at a bias voltage of -35 *V*) on sensors from the DTC-2 batch.

| Sensor type | Threshold (e) | Noise (e) |
|-------------|---------------|-----------|
| 2E | 3200 ± 58.60 | 202.3 ± 8.96 |
| 3E | 3318 ± 42.02 | 206.6 ± 8.29 |
| 4E | 3284 ± 41.27 | 229.8 ± 9.87 |

*Table 2. Threshold and noise parameters extracted from the pixel sensors from the first batch at a reverse bias voltage of 35 V [3].*

To calibrate the DUTs, radioactive γ sources (Am[241]) have been chosen. Their spectra have been compared with those obtained from measurements on ATLAS Planar Pixel Sensors. Figure 3 shows a spectrum of Am[241] acquired with a DTC-2 3E-type sensor biased at -35 V. The position of the main peak agrees with expectations within the uncertainty due to the calibration process, which was estimated to be in the order of 10-15%. Results are in good agreement with those measured with ATLAS pixel sensors [6,7].

Figure 4 shows an example of pulse height distribution of a DTC-2 3E-type detector biased at -35 *V* measured with an Sr[90] β-source. The DUT presents a most probable value (MPV) of ~ 14,070*e* with a sigma of 1,074*e* (cluster size 1) and about 15,250*e* with a sigma of 1,303*e* (cluster size 2).

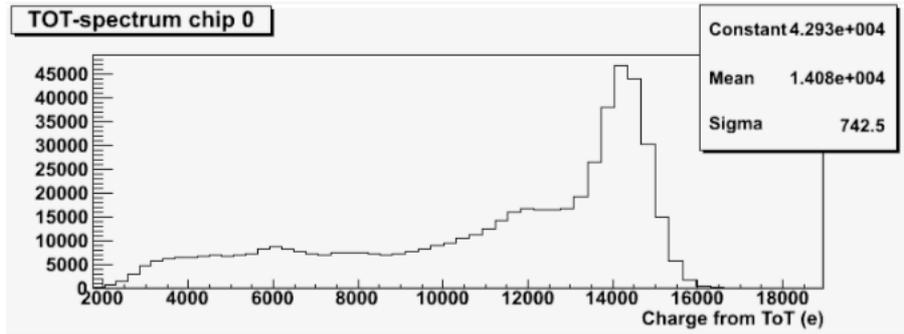

*Figure 3. Am[241] spectrum measured with DTC-2 3E detector prototype. The figure shows the source spectrum as a sum over all pixels without any clustering.*

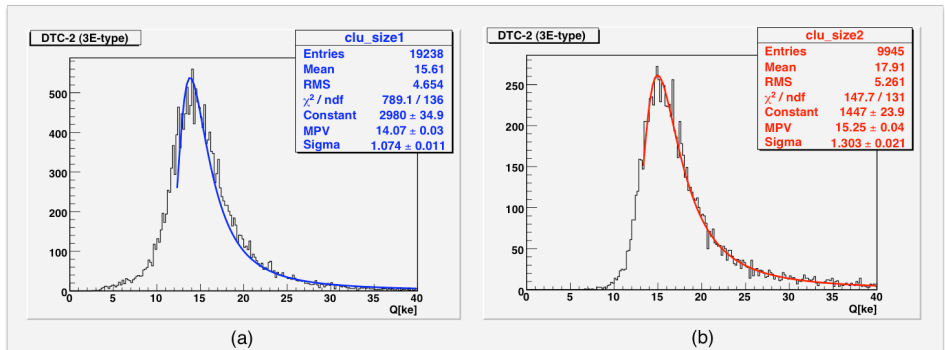

*Figure 4. Pulse height distribution for a bias voltage of -35V observed using an Sr[90] source from a DTC-2 3E sensor-type: (a) cluster size1, and (b) cluster size 2.*

Measurements with β-source have been also taken for the sensors of the second batch DTC-2B although they can be operated in a narrow bias voltage range because of the early breakdown problems. At low voltage, the noise behavior of sensors from this batch is very similar to that of sensors from the first batch.

## 4. Beam test environment

Two DDTC sensors (of 3E-type, one from each batch) have been tested in a 180*GeV/c* π[+] beam at CERN SPS with the Morpurgo super conductor dipole magnet (1.35 ± 0.10)*T* located on the H8 CERN North area beam line [8].

The aim of the beam test has been to evaluate different structures of 3D-Si sensors within a large magnetic field. Due to the orientations of the electrodes, the 3D-Si sensors are expected to be only marginally affected by the magnetic field in the ATLAS barrel configuration (E and B fields have been co-parallel) [9].

The Bonn ATLAS Telescope (BAT) has been used in this test [10]. The telescope consists of double-sided silicon micro-strip detectors with 50 μm pitched on both sides rotated by 90° with respect to each other. The trigger is given by a coincidence of two scintillators in front of and veto behind the setup.

## 5. Overview of preliminary beam tests results

### 5.1 Hit efficiency

To investigate the hit efficiency (which is obtained from tracks passing through a selected sensor region after a cuts on the quality of the track fit.) the DUTs have been irradiated in four different configurations: i) with magnetic field B on and off, and ii) with 0° and 15° angle of beam incidence [9]. Figure 5 shows the hit efficiency of the 3E-type DTC-2B (biased at -8$V$) with B field on at 0° and 15° angle of beam incidence. To show the hit efficiency as function of the track position all tracks have been studied and the track position has been folded into a 2x2 cell (800$\mu m$ x 100$\mu m$).

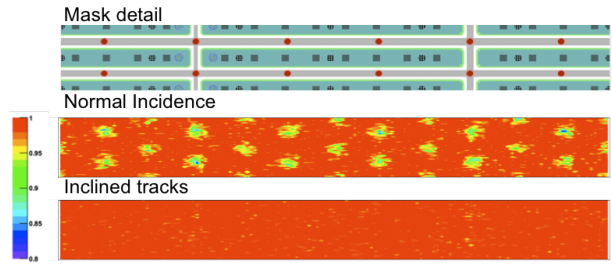

*Figure 5. Hit efficiency of the 3E-type DTC-2B with magnetic field on at 0° and 15° angle of beam incidence.*

The electrodes of both doping types are easily visible when the tracks are normal to sensor surface with a lower efficiency respect to the bulk. The hit efficiency becomes more uniform when the tracks are inclined. The overall efficiency values for sensors from both batches are summarized in Table 3.

| DUTs | B off | | B on | |
|---|---|---|---|---|
| | $\phi = 0°$ | $\phi = 15°$ | $\phi = 0°$ | $\phi = 15°$ |
| DTC-2 (3E-type) | 98.38±0.03 | 99.82±0.01 | 98.34±0.03 | 99.49±0.01 |
| DTC-2B (3E-type) | 99.21±0.02 | 99.82±0.004 | 99.14±0.02 | 99.945±0.004 |

*Table 3. The overall efficiency (%) for DTC-2 and DTC-2B devices with binomial error.*

Figure 6 shows the pulse height charge distribution for DTC-2 (a) and DTC-2B (b) sensors with B field off/on and 0°/15° angle of incidence. From Fig. 6 it is possible to see that the magnetic field does not have relevant effects on the charge distribution. The DTC-2B sensor seems to reconstruct lower charge than the DTC-2 sensors, because it was operated with a lower bias voltage (about -8 $V$) due to its early breakdown problems [3].

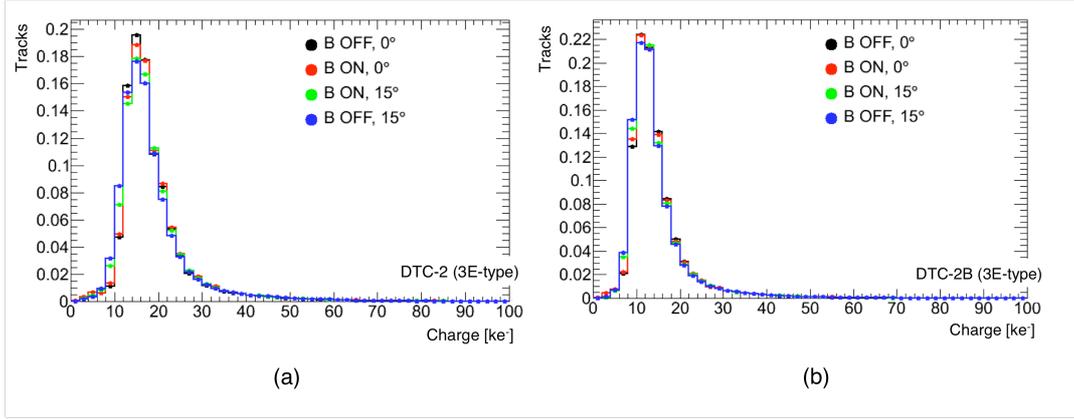

*Figure 6. Pulse height distribution for (a) DTC-2 and (b) DTC-2B sensors with magnetic field off /on and 0°/15° angle of incidence.*

## 5.2 Charge sharing

The sharing of the charge between two neighbor pixel cells has been studied defining the *charge sharing probability* as the probability that one of the relative neighbors of the pixel cell the track passes through has gone above threshold, given that the pixel cell where the track passes through has gone above threshold [9].

Table 4 summarizes the average cluster sizes for the DUTs with magnetic field off / on and at 0° /15° angle of incidence.

| DUTs | B off | | B on | |
|---|---|---|---|---|
| | $\phi = 0°$ | $\phi = 15°$ | $\phi = 0°$ | $\phi = 15°$ |
| DTC-2 (3E-type) | 1.2 | 1.7 | 1.2 | 1.6 |
| DTC-2B (3E-type) | 1.2 | 1.6 | 1.2 | 1.5 |

*Table 4. The average cluster sizes for DTC-2 and DTC-2B devices.*

Figure 7 shows (for DTC-2 3E-type) the probability of charge sharing with neighbors in the $50\mu m$ direction as function of track position within the pixel cell.

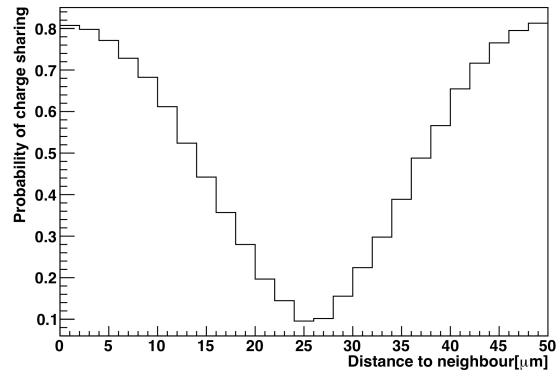

Figure 7. *Probability of charge sharing for a DTC-2 3E-type device.*

**5.3 Hit resolution**

Considering that the hit resolution improves with the amount of charge sharing, to estimate the cluster hit position the corrected charge weighted mean method has been used [9]. Figure 8 shows the residual distribution of DTC-2 device for magnetic field on and inclined tracks configuration. Table 5 presents the root mean square deviation of residual distribution for all different configurations.

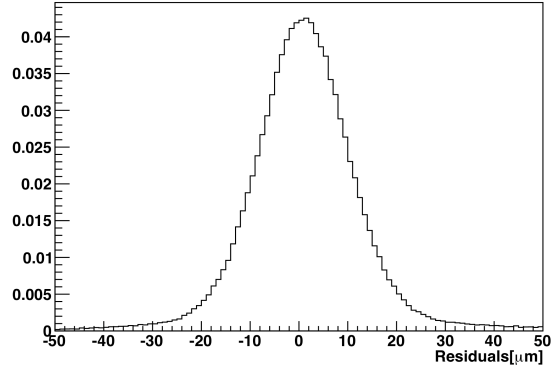

*Figure 8. Residual distribution of DTC-2 3E-type device for magnetic field on and inclined tracks configurations.*

| DUTs | B off | | B on | |
|---|---|---|---|---|
| | $\phi = 0°$ | $\phi = 15°$ | $\phi = 0°$ | $\phi = 15°$ |
| DTC-2 (3E-type) | 15.4 | 11.9 | 14.8 | 11.3 |
| DTC-2B (3E-type) | 14.0 | 10.4 | 13.5 | 9.7 |

*Table 5. The root mean square deviation of residual distribution for the four different configurations. All values are in µm.*

# 6. Conclusions

3D Silicon sensors fabricated at FBK-irst with the Double-side Double Type Column (DDTC) have been successfully tested in laboratory and in a 1.35 Tesla magnetic field with a 180 *GeV* pion beam at CERN SPS. Despite the fact that these sensors are still not optimized, in that the column overlap is still quite small compared to the substrate thickness, experimental results are satisfactory, showing good figures in terms of hit efficiency and spatial resolution and confirming that sensor performance are practically insensitive to the magnetic field. The loss of efficiency due to the empty electrodes can be largely compensated by tilting the sensors at 15° angle, this also increasing the charge sharing effects which can be exploited to improve the spatial resolution.

Further tests have already been planned to assess the behavior of these assemblies after irradiation to SLHC hadrons fluences.

**Acknowledgement**

This work ha been supported in part by the Provincia Autonoma di Trento and in part by the INFN: CN5, TREDI Project, CSN1, ATLAS Project.

The authors would like to thank: G. Gariano, A. Rovani and E. Ruscino (INFN Genova), for their precious help in system assembly and measurements; R. Beccherle (INFN Genova) for designing bump bonding mask, and S. Di Gioia (Selex SI) for the bump bonding process; H. Wilkens, the ATLAS beam tests coordinator and the CERN SPS staff for their help during data taking and installation.

**References**

[1] F. Gianotti et al. *Physic potetential and experimental challenges of the LHC luminosity upgrade*. Eur. Phys. J., vol C39 (2005) 293.

[2] A. Zoboli et al. *Double-Sided, Double-Type-Column 3-D Detectors: Design, Fabrication, and Technology Evaluation* IEE Trans. Nucl. Sci. NS-55 (5) (2008) 2275.

[3] G.F. Dalla Betta et al. *Development of 3D-DDTC pixel detectors for the ATLAS upgrade*. Paper submitted to Nucl. Instr. and Meth. A.

[4] I. Peric, et al. *The FEI3 readout chip for the ATLAS pixel Detector*. Nucl. Instr. and Meth. A 565 (2006) 178-187.

[5] SELEX Sistemi Integrati, Roma, Italy.

[6] G. Aad et al. *ATLAS pixel detector electronics and sensors*. JINST 3 (2008) P07007.

[7] F. Hugging et al. *Front-End electronics and integration of ATLAS pixel modules.* Nucl.Instr. and Meth. A 549 (2005) 157.

[8] M. Morpurgo. *A large superconducting dipole cooled by force circulation of two phase helium*. Cryogenic (July, 1979) 411.

[9] H. Gjersdal et al. *Tracking Efficiency and Charge Sharing of 3D Silicon Sensors at Different Angle in a 1.4 Tesla Magnetic Field*. Paper submitted to Nucl. Instr. and Meth. A.

[10] J. Treis et al. *A modular PC based silicon microstrip beam telescope with high speed data acquisition*. Nucl. Instr. and Meth. A 490 (2002) 112.

[11] O. Rohne et al. *Radiation Hard 3D Pixel Sensors – Recent Results*. Submitted to Vertex-2009 Proceeding in PoS.